\documentclass{aa}

\usepackage{graphicx}
\graphicspath{ {./images/} }
\usepackage{txfonts}
\usepackage[dvipsnames]{xcolor}
\usepackage{savesym}
\usepackage{longtable}
\usepackage[utf8]{inputenc}
\usepackage[T1]{fontenc}
\usepackage[]{natbib}
\usepackage{lmodern} 
\usepackage{float}
\usepackage{tabularx}
\usepackage{color}
\usepackage{ulem}
\usepackage{booktabs}
\usepackage{mathabx}
\savesymbol{second}
\savesymbol{degree}
\usepackage{multirow}
\usepackage{placeins}
\usepackage{url}
\usepackage[squaren, Gray, cdot]{SIunits}
\restoresymbol{SInits}{second}
\restoresymbol{SInits}{degree}
\usepackage[colorlinks=true,urlcolor=blue,linkcolor=blue,citecolor=blue,breaklinks=true]{hyperref}
\usepackage{appendix}
\makeatletter

\usepackage{natbib,twoopt}

\begin{document}

   \title{A search for transiting planets around hot subdwarfs}
   \subtitle{II. Supplementary methods and results from TESS Cycle 1}

   \author{A. Thuillier\inst{1,2}\and
          V. Van Grootel\inst{1}\and
          M. Dévora-Pajares\inst{3}\and
          F.J. Pozuelos\inst{1, 4}\and
          S. Charpinet\inst{5}\and
          L. Siess\inst{2}
          }

   \institute
   {
        Space sciences, Technologies and Astrophysics Research (STAR) Institute, Université de Liège, 19C Allée du 6 Août, B-4000 Liège, Belgium\\
        \email{antoine.thuillier@uliege.be or antoine.thuillier@ulb.be}
        \and
        Institut d'Astronomie et d'Astrophysique, Université Libre de Bruxelles (ULB), CP 226 1050 Bruxelles, Belgium
        \and
        Dpto. Física Teórica y del Cosmos, Universidad de Granada, 18071, Granada, Spain
        \and
        Astrobiology Research Unit, Université de Liège, Allée du 6 Août 19C, B-4000 Liège, Belgium
        \and
        Institut de Recherche en Astrophysique et Planétologie, CNRS, Université de Toulouse, CNES, 14 avenue Edouard Belin, F-31400 Toulouse, France
    }
    
\offprints{A. Thuillier}    

   \date{Received 15 March 2022 / accepted 25 May 2022}
 
  \abstract
   {Hot subdwarfs, which are hot and small He-burning objects, are ideal targets for exploring the evolution of planetary systems after the red giant branch (RGB). Thus far, no planets have been confirmed around them, and no systematic survey to find planets has been carried out.}
   {In this project, we aim to perform a systematic transit survey in all light curves of hot subdwarfs from space-based telescopes (Kepler, K2, TESS, and CHEOPS). The goal is to compute meaningful statistics on two points: firstly, the occurrence rates of planets around hot subdwarfs, and secondly, the probability of survival for close-in planets engulfed during the RGB phase of their host. This paper focuses on the analysis of the observations carried out during cycle 1 of the TESS mission.}
   {We used our specifically designed pipeline {\fontfamily{pcr}\selectfont SHERLOCK} to search for transits in the available light curves. When a signal is detected, it is processed in the next evaluating stages before an object is qualified for follow-up observations and in-depth analysis to determine the nature of the transiting body.}
   {We applied our method to the 792 hot subdwarfs observed during cycle 1 of TESS. While 378 interesting signals were detected in the light curves, only 26 stars were assigned for follow-up observations. We have identified a series of eclipsing binaries, transiting white dwarfs, and other types of false positives, but no planet has been confirmed thus far. A first computation of the upper limit for occurrence rates was made with the 549 targets displaying no signal.}
   {The tools and method we developed proved their efficiency in analysing the available light curves from space missions, from detecting an interesting signal to identifying a transiting planet. This will allow us to fulfil the two main goals of this project.}

   \keywords{planet-star interactions; planetary systems; stars: RGB; stars: horizontal branch; stars: subdwarfs; techniques: photometric}

\titlerunning{A transit survey to search for planets around hot subdwarfs: II.}
\authorrunning{A. Thuillier et al.}

\maketitle

\section{Introduction}
\label{Intro}
    Hot subdwarfs are evolved and compact stars coming in two main flavours, subdwarfs of type B (sdB) and subdwarfs of type O (sdO). The sdB stars have $T_{\rm eff} = $ 20 000 $-$ 40 000 K and log $g = $ 5.2 to 6.2 \citep{1994ApJ...432..351S,Green2008}, while sdO stars are hotter ($T_{\rm eff} = $ 40 000 $-$ 80 000 K), and have a wide range of surface gravities (log $g = $ 4.0 to 6.5;  \citealt{Oreiro2004, Johnson2014}). The sdB stars lie on the blue tail of the horizontal branch, the so-called extreme horizontal branch (EHB), which identifies them as core-He burning stars \citep{1986A&A...155...33H}. They have lost most of their envelope during the ascension of the first red giant branch (RGB), and they now have extremely thin residual H envelopes ($M_{\rm env} < 0.01 M_{\odot}$, \citealt{1986A&A...155...33H}).  This extremely thin envelope explains the atmospheric parameters of sdB stars and their inability to sustain H-shell burning, and prevents them from ascending the asymptotic giant branch (AGB) after core-He exhaustion \citep{1993ApJ...419..596D}. They rather directly evolve to the white dwarf stage through the sdO type during the immediate post-EHB phase. 
    
    In addition to the direct progenies of sdB stars, the compact sdO stars (log $g = 5.2-6.5$) could also be direct post-RGB objects through a so-called late hot He-flash \citep{2008A&A...491..253M}, or they might be end products of merger events \citep{1984ApJ...277..355W, Iben_Tutukov1984, 1990ApJ...353..215I, 2000MNRAS.313..671S, 2002MNRAS.333..121S}. Accordingly, these stars do not descend from the sdB stars.  sdO stars with log g < 5.2 exist as well, which are post-AGB stars, that is, stars that have ascended the giant branch a second time after core-He burning exhaustion \citep{2016A&A...587A.101R}. We are not interested in these in this study because we are focusing on post-RGB hot subdwarfs. An in-depth review of hot subdwarfs can be found in \citet{2016PASP..128h2001H}.
            \begin{table*}[ht!]
            \caption{\label{Sample_sectors} Detailed statistics for hot subdwarfs observed in SC mode during the primary mission of TESS (July 2018 -- July 2020). Similar to Table 1 from \citet{Van-Grootel-21}, but with details for both cycles.}
                \begin{center}
                    \begin{tabular}{|c c c c|}
                    \hline\hline
                    \textbf{Number of} &\textbf{Primary}& \textbf{Cycle} & \textbf{Cycle} \tabularnewline
                    \textbf{Sectors} & \textbf{mission} & \textbf{1} & \textbf{2} \tabularnewline
                    \hline
                     1 & 877 & 627 & 250 \tabularnewline
                     2 & 205 &  95 & 110 \tabularnewline
                     3 &  72 &  25 &  47 \tabularnewline
                     4 &  23 &   7 &  16 \tabularnewline
                     5 &  21 &   3 &  18 \tabularnewline
                     6 &  24 &  10 &  14 \tabularnewline
                     7 &   7 &   2 &   5 \tabularnewline
                     8 &  10 &   5 &   5 \tabularnewline
                     9 &   6 &   3 &   3 \tabularnewline
                    10 &   6 &   1 &   5 \tabularnewline
                    11 &  13 &   3 &  10 \tabularnewline
                    12 &  23 &   7 &  16 \tabularnewline
                    13 &  15 &   4 &  11 \tabularnewline
                    \hline
                    \textbf{Total} & 1302 & 792 & 510\tabularnewline
                    \hline
                    \textbf{Mean sect./star} & 2.1 & 1.6 & 2.8\tabularnewline
                    \hline \hline
                \end{tabular}
            \end{center}
        \end{table*} 

    There are several planet candidates around hot subdwarfs. They were identified through various methods such as reflection signals in Kepler light curves \citep{2011Natur.480..496C, Bear2014} or stellar pulsation timing variations \citep{2007Natur.449..189S}. A review of the search for planets around hot subdwarfs can be found in \citet{Van-Grootel-21}. None of these candidates have been confirmed to date; some are heavily debated \citep{2015A&A...581A...7K, 2019A&A...627A..86B}, and others were discarded later \citep{2018A&A...611A..85S, Krzesinski2020IAU}. A mini radial velocity (RV) survey carried out with the HARPS-N spectrograph of eight apparently single hot subdwarfs gave null results down to a few Jupiter masses \citep{2020arXiv200204545S}. More generally, in regard to horizontal branch (HB) stars, a single planet around an F2-type HB star was announced by \citet{2010Sci...330.1642S}, but was later discarded by \citet{2014A&A...562A.129J}. The transit method has never been used at a large scale to search for planets transiting hot subdwarfs. 
    
    Due to the loss of most of their envelope during the RGB, hot subdwarfs are small stars ($0.1$ to $0.3$ $R_{\odot}$; \citealt{2016PASP..128h2001H}). This property makes them ideal targets for using the transit method to address the question of the evolution of exoplanetary systems directly after the RGB phase of evolution. We indeed aim in this project to perform a transit survey in all available light curves of hot subdwarfs from space-based telescopes. The main objective is to determine the occurrence rates of planets around these stars as a function of orbital period and planetary radius. We also aim to place strong observational constraints on the survival of close-in planets that were engulfed during the RGB phase of their host star, which are currently completely missing. "Close-in planets" here mean those with current orbital periods up to $\sim$ 50 days (orbital radii up to 0.20 AU), which have the highest transit probability (which is about 0.35\% at 50 d orbital period). Considering main-sequence masses between 1 and $\sim$2.5 $M_{\odot}$ that are the main hot subdwarf progenitors, all these planets could have been engulfed during the RGB phase of the host star \citep{2009ApJ...705L..81V}. 
    
    This paper is the second of the series started by \citet{Van-Grootel-21}, where the context of the project was introduced, together with the results from injection-and-recovery tests on actual light curves from the Kepler \citep{2010Sci...327..977B}, K2 \citep{2014PASP..126..398H}, and TESS \citep{2014SPIE.9143E..20R} space missions. From these performance tests, we determined  which transiting bodies in terms of object radius and orbital period we are able to detect with our tools. In TESS data, we are able to detect transiting bodies of $\lesssim 2 R_{\oplus}$ with orbital periods shorter than 15 days for a magnitude (G mag) between 13 and 14 and a single sector (27 days) of observations \citep{Van-Grootel-21}. For brighter stars and/or stars that were observed in more sectors (up to 13 for the primary mission, which represents one year of continuous observation), we are able to detect bodies with radii smaller than $1 R_{\oplus}$, as well as bodies with radii of a few $R_{\oplus}$ for orbital periods longer than 35 days (see details in Table 3 of \citealt{Van-Grootel-21}). 

    This second paper focuses on the full explanation of the light-curve analysis process and on the first results obtained for the hot subdwarfs observed by TESS during its cycle 1. 
    Section \ref{datamethod} presents the data we used and the sample selection process, followed by a description of the method we used for the transit survey along with a brief description of our main tools. We also detail the different steps of the analysis and how we ranked the detected signals. In Sect. \ref{results} we present the results and detail the number of signals in each stage of the analysis. We also report some examples of false positives that we encountered during our survey. We conclude in Sect. \ref{discussion} with a discussion and describe the next steps of the project.

\section{Data and method}
    \label{datamethod}
    \subsection{Data}

        TESS observed 1302 sdB and sdO stars in 2-minute short-cadence (SC) mode during its primary mission \citep{Van-Grootel-21}. Two so-called cycles compose this primary mission. Cycle 1 (July 2018 -- July 2019) observed the southern and cycle 2 (July 2019 -- July 2020) the northern celestial hemisphere. The two hemispheres are revisited during the extended mission (July 2020 -- July 2022), with cycle 3 (southern) and cycle 4 (northern). Except for cycle 4, which has 16 sectors, each cycle is divided into 13 sectors of $\sim$27 days of observations, corresponding to two orbits of the satellite. 

        There is a significant disparity in the length of the observations available from target to target. The TESS sectors overlap near the celestial poles, which means that a few targets have been observed almost continuously during one year, while the large majority are visible in a single sector of only $\sim$27 days. Table \ref{Sample_sectors} displays the detailed statistics for hot subdwarfs for cycles 1 and 2 of the primary mission. Interestingly, a difference appears between the two cycles: the number of stars in one single sector in cycle 1 is significantly larger than in cycle 2.  This is striking when the mean number of sectors per star is computed, which is 1.6 in cycle 1 and 2.8 in cycle 2. This discrepancy is directly linked to the target positions in the sky (see Fig.~1 from \citealt{Van-Grootel-21}), which in turn is most likely linked to our position in the Galaxy and our height above the Galactic plane.

        We focus in this paper on the hot subdwarfs observed in TESS cycle 1, completed by cycle 3 in case of detection of an interesting signal (see below). Cycle 1 contains sectors 1 to 13, while cycle 3 contains sectors 27 to 39. Cycle 2 (sectors 14 to 26) and cycle 4 (sectors 40 to 56) will be explored in the next steps of this project. TESS provides flux measurements with different exposure times. The work presented here mainly used the 2-minute SC mode. In its primary mission, TESS only provided data with cadences of 2 and 30 minutes, while in its extended mission, two additioanl observation modes are available at 20-second and 10-minute cadences. We occasionally used these other cadences (20 seconds, 10 minutes, and 30 minutes) to probe interesting signals detected in the 2-minute mode.

        The data used in this study are the pre-search data conditioning simple aperture photometry (PDC-SAP) light curves, which are the sum of the flux from the pixels in a pre-defined aperture corrected for long-term trends. They were directly downloaded from the NASA Mikulski Archive for Space Telescopes (MAST) database. Other data were gathered from the TESS 
        Asteroseismic Science Operation Center, for example the identification of the pulsating hot subdwarfs.
        
    \subsection{Light-curve analysis: Method and tools}
        \subsubsection{SHERLOCK pipeline}
            The main analysis was conducted using our pipeline Searching for Hints of Exoplanets fRom Light curves Of spaCe-based seeKers {\fontfamily{pcr}\selectfont SHERLOCK}\footnote{SHERLOCK's code is open-source, user-friendly, and available on GitHub: \url{https://github.com/franpoz/SHERLOCK}} \citep{Pozuelos-20}, which is a versatile tool we recently developed to detect shallow periodic transits in light curves from space-based observatories such as Kepler/K2 and TESS. {\fontfamily{pcr}\selectfont SHERLOCK} has six different modules that allow (1) downloading and preparing the light curves from their online repositories, (2) searching for planetary candidates, (3) performing a semi-automatic vetting of the interesting signals, (4) computing a statistical validation, (5) modeling the signals to refine their ephemerides, and (6) computing observational windows from ground-based observatories to trigger a follow-up campaign. In addition, to optimize the planetary search, {\fontfamily{pcr}\selectfont SHERLOCK} executes an automatic process that allows for initial corrections such as masking high-noise regions and correcting strong variability caused by fast rotators. Then, {\fontfamily{pcr}\selectfont SHERLOCK} performs a user-defined multi-detrend approach using the bi-weight method \citep{wotan} by varying the window size a number of times. In our case, we used 12 different detrends. Then, {\fontfamily{pcr}\selectfont SHERLOCK} searches for planetary candidates in the original PDC-SAP flux jointly with the 12 newly detrended light curves. This strategy allows us to search for the most appropriate detrend maximising the signal-to-noise ratio (S/N) and  signal detection efficiency (SDE)\footnote{SDE is computed as $\frac{1- \langle SR \rangle}{\sigma(SR)}$ , with \textit{SR} defined as the signal residue for a tested period, and $\sigma$ is the standard deviation.} of any detection. The six modules are implemented so that the user only needs to execute a few lines of code, which avoides diving into many different pipelines, codes, and processes. Moreover, {\fontfamily{pcr}\selectfont SHERLOCK} has direct access to short- and long-cadence data observed by Kepler/K2 and TESS.
            
\begin{figure}[ht!]
                \includegraphics[width=9cm]{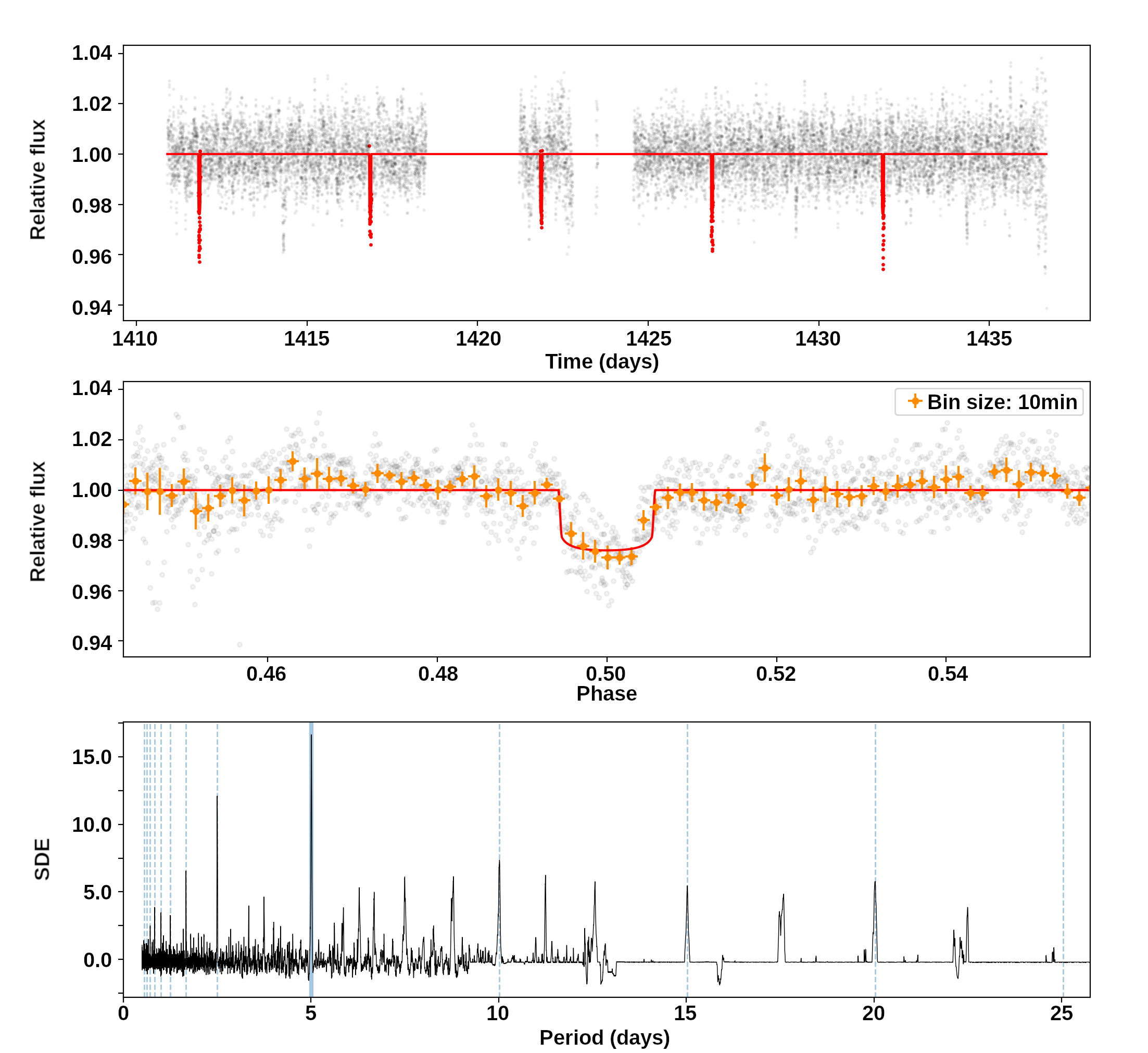}
                \includegraphics[width=9cm]{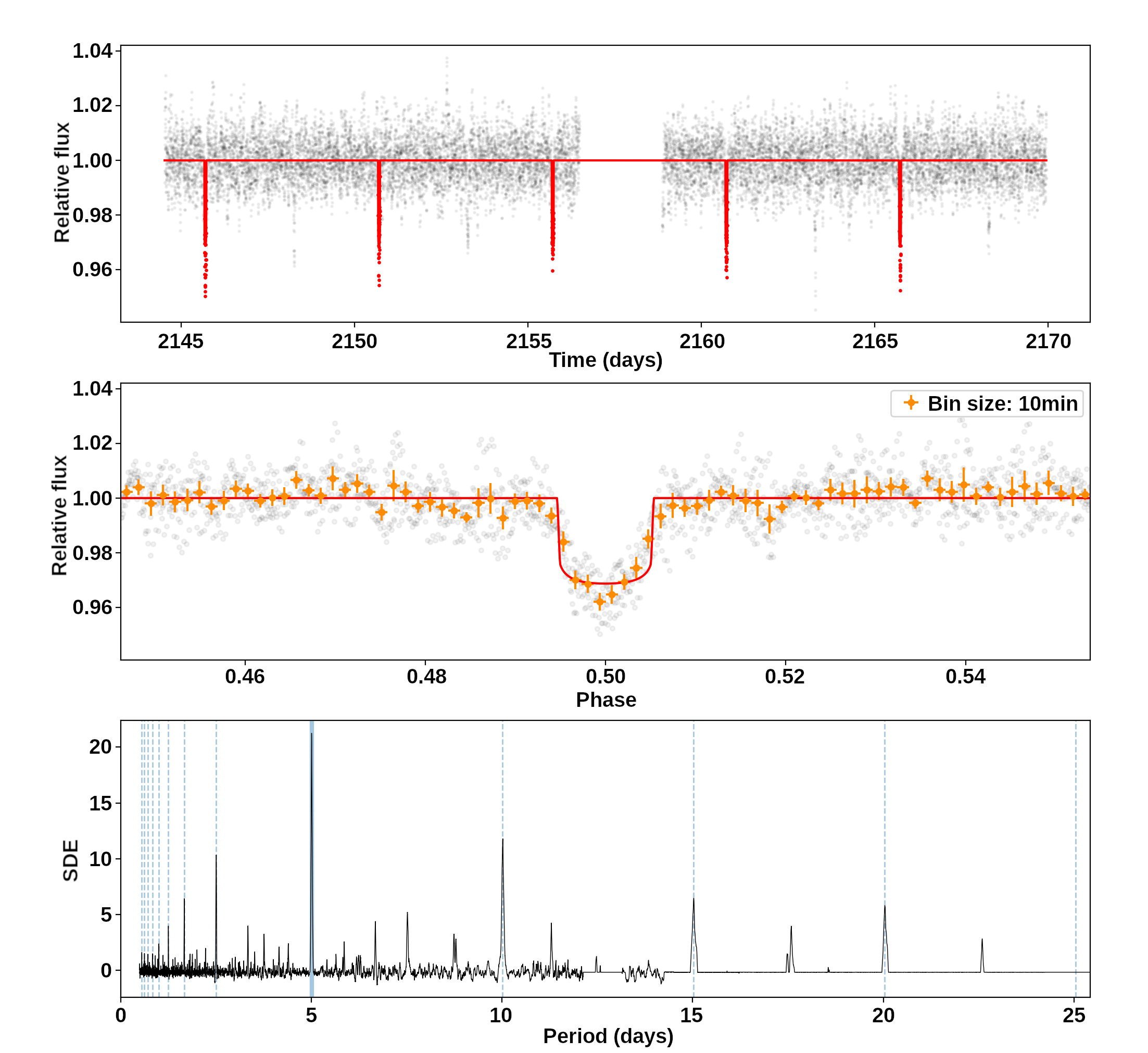}
                \caption{\label{outputSherlock}{\fontfamily{pcr}\selectfont SHERLOCK} output of the best result of the first run of TIC 142875987 in sectors 4 (panels 1 to 3, from top to bottom) and 31 (panel 4 to 6) of TESS. The secondary eclipse is clearly visible in the top panel of both sectors (panels 1 and 4) where the detrended light curve (grey) and position of the detected transits (red) are shown. Panels 2 and 5: Phase-folded light curve over the period of the spotted transit. Panels 3 and 6: Power spectrum with the main signal and its harmonics highlighted.}
                \centering
            \end{figure}

            A typical {\fontfamily{pcr}\selectfont SHERLOCK} execution starts with the automatic download of the PDC-SAP flux light curve of the desired target from the MAST. Depending on the options that are set, it is possible to analyse all the available data of the given target or use only specified sectors (for TESS), quarters (Kepler), or campaigns (K2). Then, after the initial corrections described above, the planetary search starts using a modified version of the transit least- squares (TLS) package \citep{tls}. This modified version consists of a proper reduction of the period-range density over which TLS searches for planets. This reduction allows us to avoid extremely dense searches due to observations performed over sectors with large time gaps between them, which translates into a much faster but less sensitive execution. When {\fontfamily{pcr}\selectfont SHERLOCK} spots a periodic signal with S/N and SDE above the minimum thresholds defined by the user, in our case, 6 and 8, respectively, this signal is masked, and the search is repeated until no other signals above the thresholds are found. This search-and-mask process is called ``run''. In this study, we set a limit of six runs to avoid wasting computational time while maximising the chances to detect transits. After six runs, the light curves have many masked regions coming from the previous findings, implying that any new detection will be affected by these gaps. This reduces their credibility and therefore makes the following validation checks more challenging. All the information regarding the applied detrending, initial masking, fast rotation correction, and planetary searches are stored automatically in folders and log files. Then, the user needs to inspect the results to verify the findings visually. An example of a {\fontfamily{pcr}\selectfont SHERLOCK} finding is shown in Fig.~\ref{outputSherlock}. The figure shows two different observations of TIC 142875987, one during cycle 1 (top three panels), and the other from cycle 3 (bottom three panels). Panels (1) and (4) (from top to bottom) show the light curve with the position of the spotted transits, panels (2) and (5) display the phase light curve folded over the period of the detected signal, while panels (3) and (6) show the power spectrum of the phase-folded light curve for all the possible periods, with highlights on the selected signal and its harmonics.
            
        \subsubsection{FELIX}  
            \begin{figure*}[ht!]
                \label{fig:FELIX}
                \includegraphics[scale=0.25]{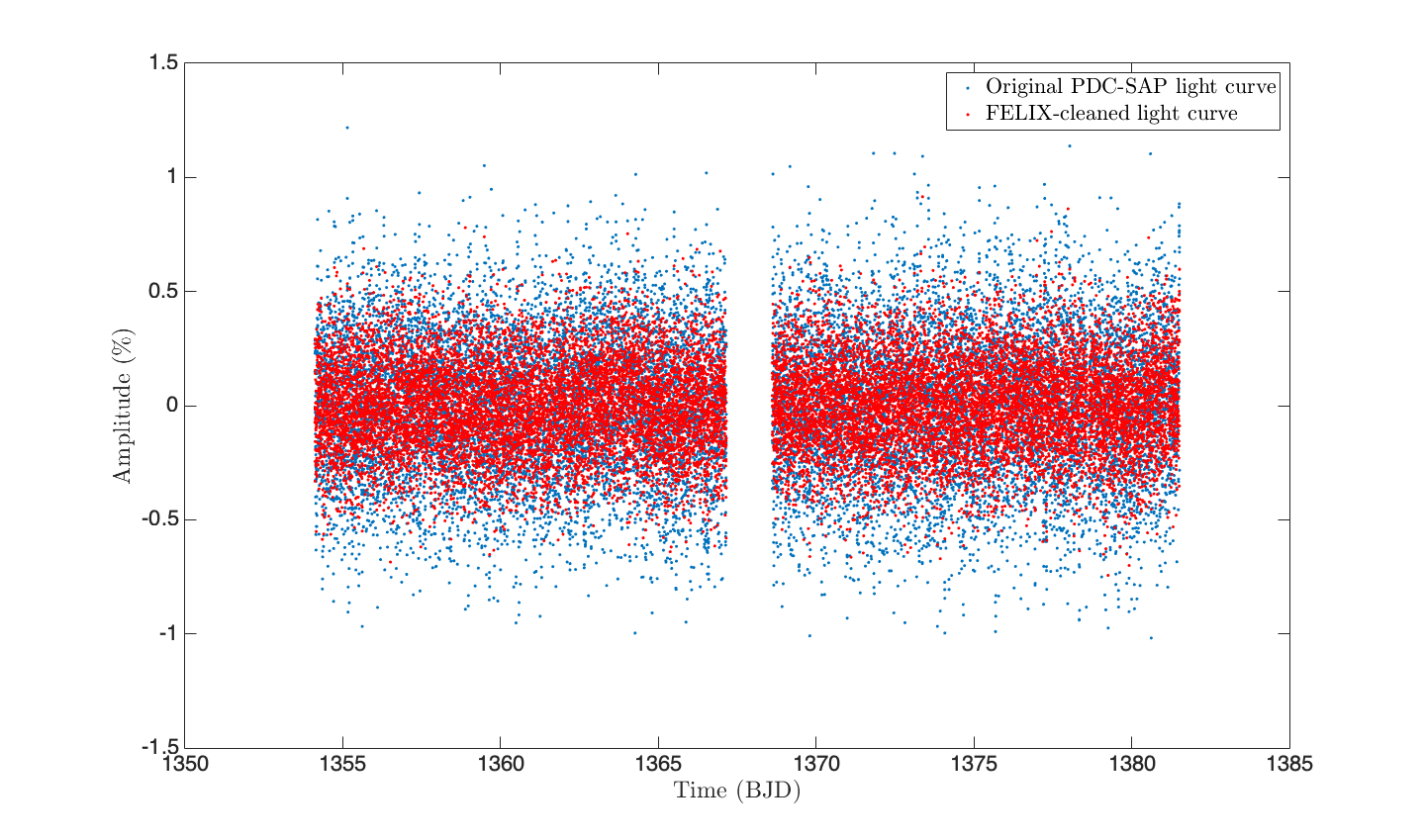}
                \includegraphics[scale=0.45]{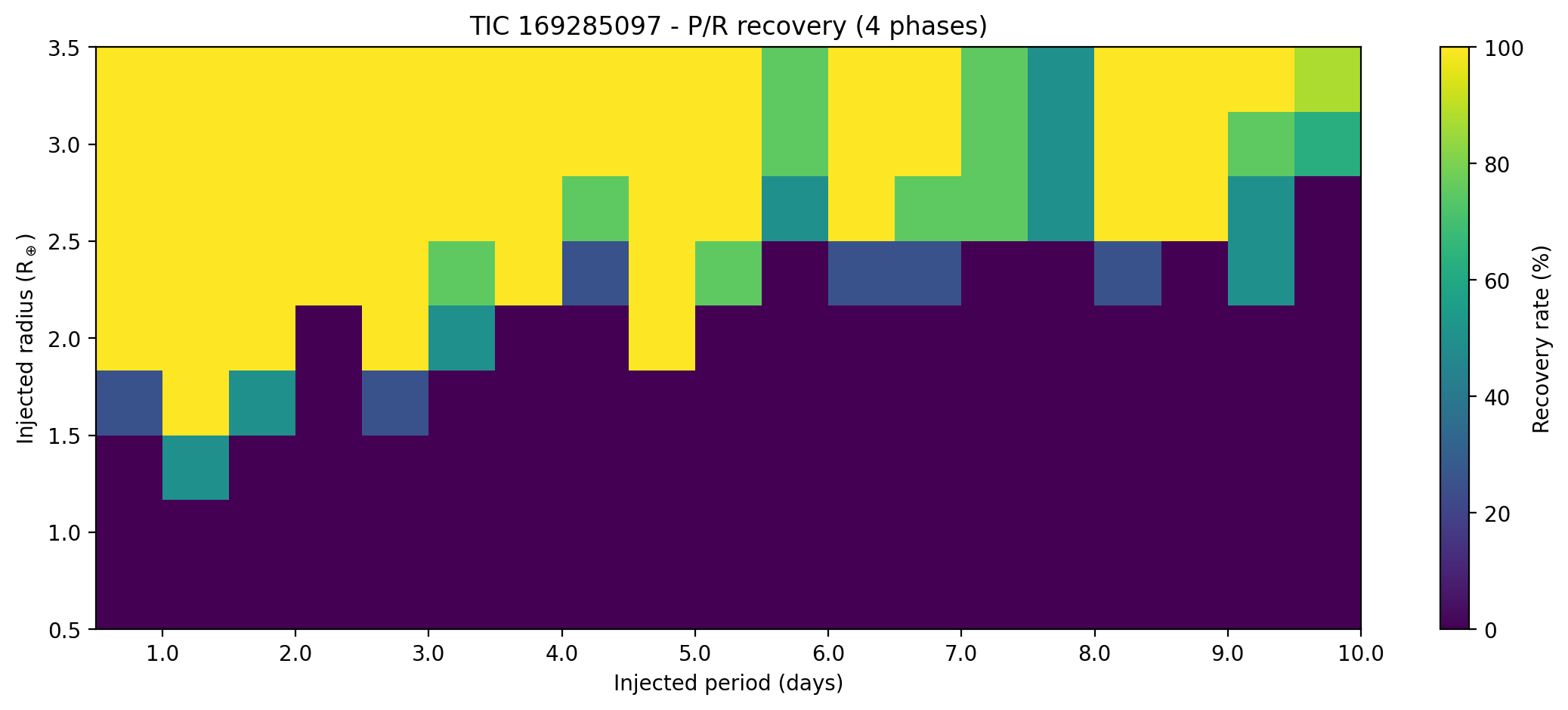}
                \includegraphics[scale=0.45]{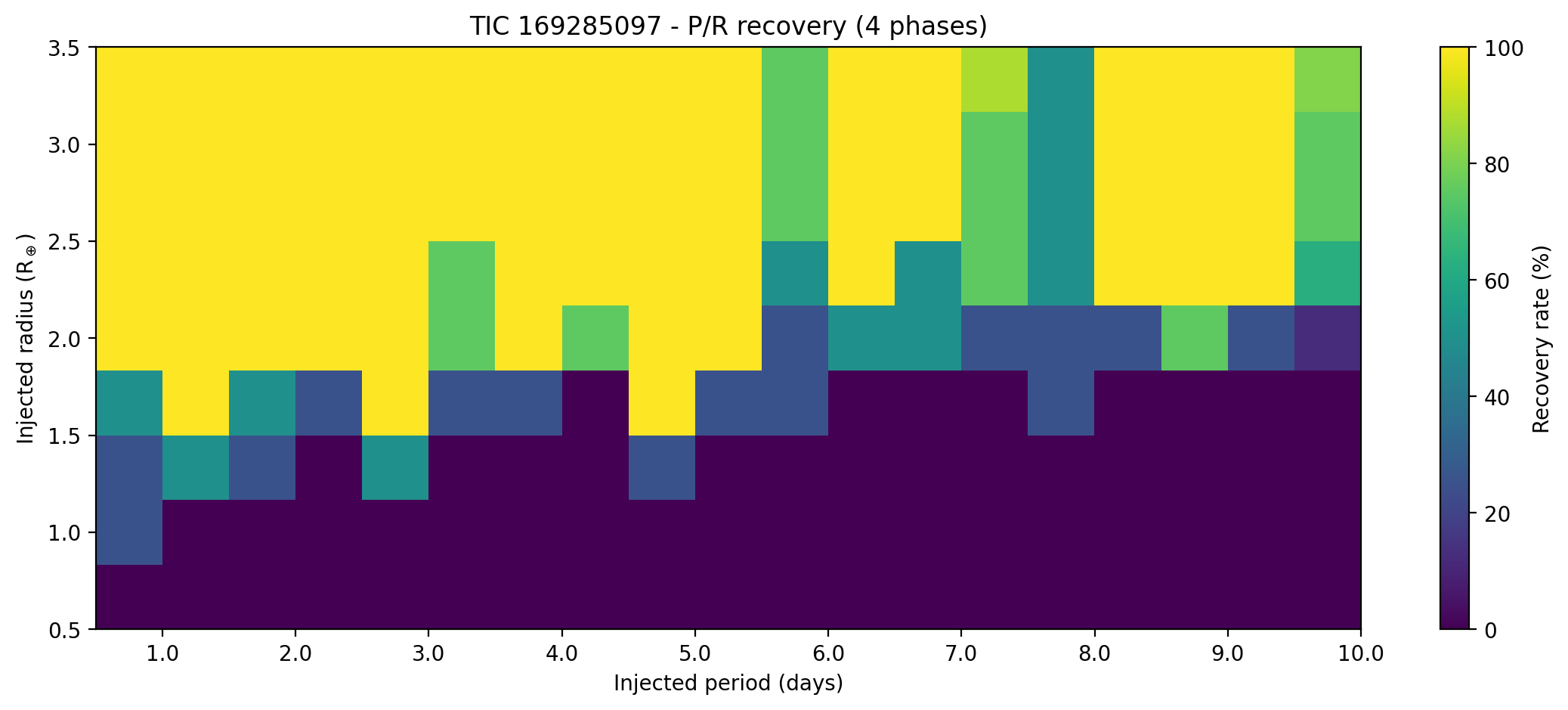}
                \centering
                \caption{sdB hybrid pulsator TIC 169285097 (Gmag=10.92) observed in Sector 2 by TESS. Panel 1 (top): Original (DQUALITY=0 datapoints) PDC-SAP light curve (blue) and light curve after the stellar oscillations were removed (red). Panel 2 (middle): Injection-and-recovery test on the original light curve. Panel 3 (bottom): Injection-and-recovery test on the processed light curve by FELIX. In panels 2 and 3, injected planets have radii in the range of 0.5–3.5 ~$R_{\Earth}$ (steps of 0.37~$R_{\Earth}$) and orbital periods of 0.5–10 d (steps of 0.2~d). For each combination of radius and orbital period, we analysed four epochs (or phases), that is, each panel evaluates 760 scenarios computed with our tool {\fontfamily{pcr}\selectfont MATRIX}. Drops in detection at $\sim$6, 8, and 10 days in period are target induced, probably due to noise correlation at these precise periods.}
            \end{figure*}
            
            Some of the light curves display too much variability, typically related to stellar oscillations or to the presence of a companion star, to be directly analysed by the {\fontfamily{pcr}\selectfont SHERLOCK} pipeline. In these cases, we used FELIX \citep{2010A&A...516L...6C} \citep{Zong2016}, a tool designed to extract interactively or automatically periodic variations in a light curve. FELIX subtracts from the light curve each periodic variation that is spotted above a pre-defined threshold (usually the threshold corresponding to a 4$\sigma$ significance) using the pre-withening technique \citep{Deeming-75}. That is, we identified the frequency and amplitude of the highest-amplitude peak in the Lomb-Scargle Periodogram (LSP) of the light curve. They were used as initial guesses in a subsequent non-linear least-squares (NLLS) fit of a cosine wave in time domain using the Levenberg-Marquardt algorithm. The fitted wave of the derived frequency, amplitude, and phase was then subtracted from the light curve. The operation was repeated as long as there was a peak above the pre-defined threshold. The light curve cleaned from stellar oscillations and other periodic variabilities (e.g. linked to binarity) was then given to {\fontfamily{pcr}\selectfont SHERLOCK} to start the transit search. This cleaning by FELIX improved our capability to detect small planets in light curves dominated by stellar oscillations or other periodic variabilities. An example is given in Fig.~\ref{fig:FELIX}, which presents the case of the hybrid sdB pulsator TIC 169285097. The upper panel presents the light curve (SC cadence, sector 2) before (original) and after the pre-withening process, that is, after the oscillations were removed. The two bottom panels show the results of injection-and-recovery tests performed with our Multi-phAse Transits Recovery from Injected eXoplanets {\fontfamily{pcr}\selectfont MATRIX}\footnote{{\fontfamily{pcr}\selectfont MATRIX} is open access at https://github.com/PlanetHunters/tkmatrix} tool \citep{Matrix2022}, on the original and cleaned light curves, respectively. These tests were made following the same procedure as described in \citet{Van-Grootel-21}. The results provided by the inject-and-recovery tests show the benefits of using FELIX. In panel 2, we might not detect planets with radius $\lesssim$1.5~$R_{\Earth}$ for short orbital periods $\lesssim$3.0~days, and hardly detect planets with $\lesssim$2.0--2.5~$R_{\Earth}$ with orbital periods $\gtrsim$3.0~days. However, in panel 3, after the cleaning performed by FELIX, our detection limits improved. In this situation, we started to detect planets with a radius ranging from 1.0-1.5~$R_{\Earth}$ for short orbital periods $\lesssim$3.0~days, and $\sim$1.5--2.0~$R_{\Earth}$ for longer orbital periods. 
            
            We explicitly confirmed, again by injection-and-recovery tests, that no transit was removed from the light curve by our pre-withening procedure. Transits translate into an LSP by a comb of frequencies with decreasing amplitudes, with the orbital frequency $f_{\rm orb}$ (the highest-amplitude peak) and its harmonics $n*f_{\rm orb}$. The number $n$ of harmonics detected depends on the transit depth and on the number of transits in the light curve. In rare cases, $f_{\rm orb}$ might be the only peak detected above our usual threshold of 4$\sigma$ significance. If this orbital frequency is of the order of g-mode pulsation frequencies ($\sim$30 min to 3 hr) in an identified g-mode pulsator, the $f_{\rm orb}$ peak, and hence the transit signal, might be removed in our pre-withening procedure because it would be misidentified as a g-mode pulsation. This situation is quite rare, however, and never occurred during our extensive tests.
            
            
        \subsubsection{Visual inspection}
\begin{figure}[ht]
                \includegraphics[width=9cm]{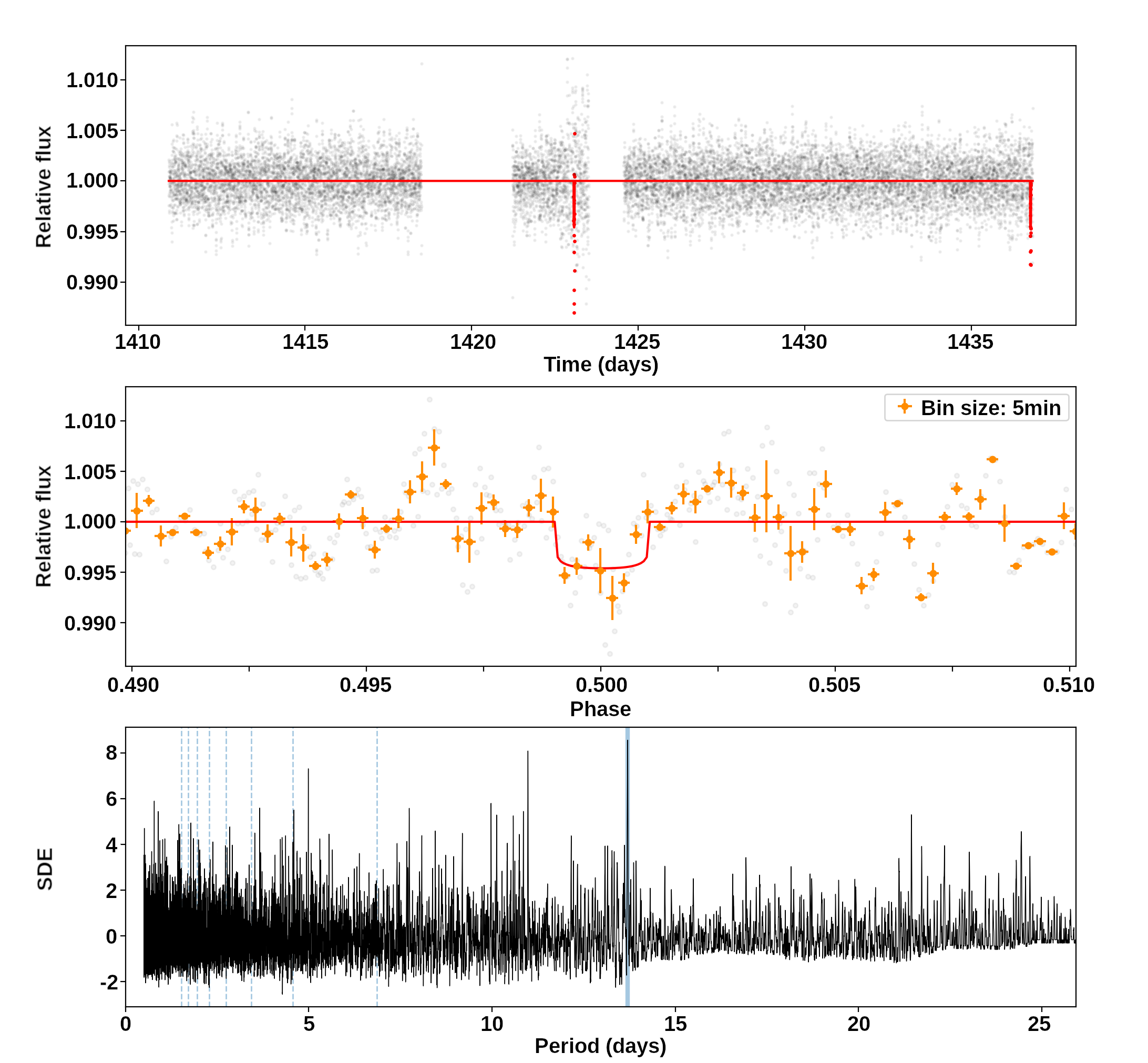}
                \caption{\label{fig:visual-negative} Example of a signal in the light curve of TIC 64111698 with both S/N and SDE above our thresholds. The light curve was still ruled out during our visual inspection. The position of the two detected transits is suspicious (top panel): the first transit occurs is in a high-noise region, and the second transit is at the edge of a dataset, where typically more trends exist. The phase-folded light curve (middle panel) shows strong variability out of transit with a similar amplitude as the transit signal. Finally, the signal has no harmonics and does not significantly rise above the noise (bottom panel). All these hints suggest that the spotted signal is most likely a false positive.}
                \centering
            \end{figure}

            When a {\fontfamily{pcr}\selectfont SHERLOCK} execution was completed, a visual analysis of all the signals satisfying the S/N and SDE thresholds was performed to prove their credibility. Artefacts in the light curve were frequently mistaken as transits. This step allowed us to confirm or remove this from the list of interesting signals. One example of this visual invalidation is shown in Fig. \ref{fig:visual-negative}. When a signal successfully passed the visual inspection, an analysis of cycle 3 data -- if available\footnote{Only 12 hot subdwarfs out of the 792 were not re-observed during cycle 3.} -- from the same target was carried out with the same conditions. The result was then carefully studied to detect any signal that matched the one spotted in cycle 1 (in this phase, signals slightly below thresholds are accounted for). If none was detected, the target was set aside and  awaited potential new observations. If the signal was confirmed, the target was worth an individual in-depth investigation and proceeded the next steps, which were a literature check. The vetting process was triggered as well.
            
        \subsubsection{Literature check}
            Databases such as ExoFOP\footnote{ExoFOP: Exoplanet Follow-up Observing Program, \url{https://exofop.ipac.caltech.edu/tess/}} and SIMBAD\footnote{SIMBAD: Set of Identifications, Measurements and Bibliography for Astronomical Data, \url{http://simbad.u-strasbg.fr/simbad/}} were searched for a previous detection with similar properties, and we reviewed the papers that studied each individual target. If a signal matching our results was reported previously and a planetary origin excluded by the data, the case was closed. This was the case of TIC 142875987 (Fig.~\ref{outputSherlock}):  \citet{Bell-19} determined that this transit
comes from a low-mass white dwarf closely orbiting the hot subdwarf.
            
        \subsubsection{Vetting process} 
            \label{vetting}
            A vetting procedure of signals satisfying the conditions above was made with {\fontfamily{pcr}\selectfont SHERLOCK} using the 
            LATTE package\footnote{LATTE: light curve analysis tool for transiting exoplanet, \url{https://github.com/noraeisner/LATTE}}. For each 
            individual transit event, the vetting includes a check of the following elements:
        
            \begin{itemize}
                \item The light curve around each transit overplotting TESS's momentum dumps.
                \item The background flux variation.
                \item Monitoring of the $x$ and $y$ positions of the brightest star in the aperture as a function of time to test the TESS stability.
                \item The aperture-size dependence.
                \item The average flux in and out of the transit.
                \item The location and brightness of nearby stars.
                \item The light curves from each individual pixel in the target pixel file in transit times.
            \end{itemize}
            
            More details about LATTE can be found in \citet{latte2020}. Then, we performed a statistical validation using the TRICERATOPS package\footnote{TRICERATOPS: A tool for vetting and validating TESS objects of interest, \url{https://github.com/stevengiacalone/triceratops}}. TRICERATOPS uses a Bayesian framework to compute the probabilities of various astrophysical transit-producing scenarios such as a transiting planet with a given orbital period (P$_{\rm orb}$), an eclipsing binary (EB) with P$_{\rm orb}$, or an EB with 2$\times$P$_{\rm orb}$ (see Table 1 in \citealt{triceratops} for all the scenarios tested). For each scenario, a false positive probability (FPP) and nearby false positive probability (NFPP) are provided, which allow us to verify the reliability of a given candidate.

        \subsubsection{Follow-up of positive signals}          
            If a given signal fulfiled the visual inspection, vetting, and statistical validation steps, an observation with high spatial resolution is required to confirm the signal on the target star. This is particularly critical for TESS candidates. The pixel scale of TESS is 21'' , and its point-spread function may be as large as 1', both of which increase the probability of contamination by a nearby EB \citep[see e.g.][]{kostov2019}. We did not directly use the CROWDNESS factor provided by the TESS team as we independently checked nearby stars at the previous stage of the analysis (vetting process). Before triggering a follow-up campaign, we first refined the transit parameters. This is because the results coming directly from the search step through the modified-TLS algorithm are not optimal. The associated uncertainties on orbital period (P$_{\rm{orb}}$), epoch (T$_{0}$), and transit duration ($d$) are too high to schedule an efficient observation. Hence, to refine the transit parameters, {\fontfamily{pcr}\selectfont SHERLOCK} used the results coming from the modified TLS as priors to perform model fitting, injecting them into \texttt{allesfitter} \citep{allesfitter-code,allesfitter-paper}. The fitting was made using the dynamic nested sampling algorithm, whose posterior distributions are much more refined, with significant reductions of a few orders of magnitude of the uncertainties on P$_{\rm{orb}}$, T$_{0}$ , and $d$. This new set of parameters allowed us to schedule a follow-up campaign with reliable observational windows. 
            
            Then, depending on the transit parameters and stellar characteristics, two options are available to our project. For transit signals whose depths exceeded 2.5~mmag and for a stellar brightness in the range of 8 to 15 G magnitude, we used the TRAPPIST\footnote{TRAPPIST: TRAnsiting Planets and PlanetesImals Small Telescope} telescope network. The TRAPPIST network is composed of two ground-based 60~cm telescopes, situated in La Silla Observatory in Chile \citep{Jehin-11} and in Ouka{\"{\i}}meden Observatory in Morocco \citep{barkaoui2019}. On the other hand, for signals shallower than 2.5~mmag and a stellar brightness $\leq$14 G magnitude, we used the CHEOPS ESA mission \citep{2020ExA...tmp...53B}\footnote{CHEOPS mission details are available on the European Space Agency website: \url{https://sci.esa.int/web/cheops/-/54030-summary}} via our observational program presented by \cite{Van-Grootel-21}.
            
            At this step, a signal that is successfully recovered by the follow-up campaign will trigger an additional investigation to characterise the nature of the transiting body. Spectroscopic analysis and spectral energy distribution (SED) fitting \citep{2018OAst...27...35H} will be carried out to constrain the stellar parameters and, in particular, its radius. A stellar, white dwarf, or brown dwarf origin for the transiting body will need to be ruled out based on radial velocity (RV) measurements. We will first search for RV data in archives that are open to the community (such as the ESO archives) or within the hot subdwarf community. If needed, we write proposals for appropriate spectrographs on a case-by-case basis.

\section{Results}
    \label{results}
    Of the 792 hot subdwarfs identified in the SC observations of cycle 1 of TESS, 549 did not have a visually credible transit signal. These stars are listed in Table \ref{Negative_list} and their celestial distribution are shown in Figure \ref{figA1}. Conversely, 243 targets displayed at least one signal requiring further investigation. Because some targets displayed several interesting signals, their total number is greater than the number of targets with positive results: altogether, we identified 352 potential signals. To keep track of the progress of the analysis of this large number of signals, they were ranked as follows:

    \begin{itemize}
        \item Stage 0: The signal has been detected above our thresholds by {\fontfamily{pcr}\selectfont SHERLOCK} and is visually credible.
        \item Stage 1: A signal with similar properties is detected in the extended mission (cycle 3 of TESS).
        \item Stage 2: The vetting confirms that it does not come from a nearby star or from unrelated background features, among other checks (see Sect. \ref{vetting}).
        \item Stage 3: Follow-up observations independently confirm the existence of the signal.
        \item Stage 4: Data confirm the planetary nature of the candidate.
    \end{itemize}
    
    \begin{table}[!ht]
        \caption{\label{Stages} Number of signals at each stage of the analysis.}
        \begin{center}
            \begin{tabular}{|c c l|}
                \hline\hline
                \textbf{Stage} & \textbf{Signals} & \textbf{Comment} \tabularnewline
                \hline
                0 & 352 & {\fontfamily{pcr}\selectfont SHERLOCK} + visual positive \tabularnewline
                1 & 46 & also spotted in Cycle 3 \tabularnewline
                2 & 30 & passed vetting + validation process
                \tabularnewline
                3 & 2 & recovered in follow-up observations \tabularnewline
                4 & 0 & planetary nature confirmed \tabularnewline
                \hline \hline
            \end{tabular}
        \end{center}
    \end{table}
    
    Of the 352 signals in stage 0, 294 were ruled out (they did not match the criterion for stage 1) and 12 belong to stars that were not re-observed in cycle 3 (not shown in Table \ref{Stages}). Therefore, only 46 went through the vetting and validation process. Thirty of them reached stage 2 and became the subject of a follow-up campaign currently active (the position in the sky of these followed-up targets is plotted in Figure \ref{figA1}). At the time of writing, 7 targets have been observed and other observations are scheduled for the coming months. No signal of confirmed planetary origin has been detected so far. However, several targets have already retained our attention, and some of them are detailed in the following paragraphs.
    
    A first example is the signal detected in the light curve of TIC 142875987 (PHL 1539), in which transits were spotted in the sector 4 (panels 1 to 3 of Fig.~\ref{outputSherlock}). The transiting body has a period of 5.01 days, and the signal is well above our thresholds (S/N 30.7, SDE 16.6, see Fig.~\ref{outputSherlock}). Therefore data from cycle 3 of this target were analysed and the detection was retrieved, as shown in panels 4 to 6 of Fig. \ref{outputSherlock}. However, the literature check revealed that this signal has been investigated by \citet{Bell-19}, and the authors determined that it is a low-mass white dwarf in an EB configuration with the targeted sdB star.
    
\begin{figure}[ht]
        \includegraphics[width=9cm]{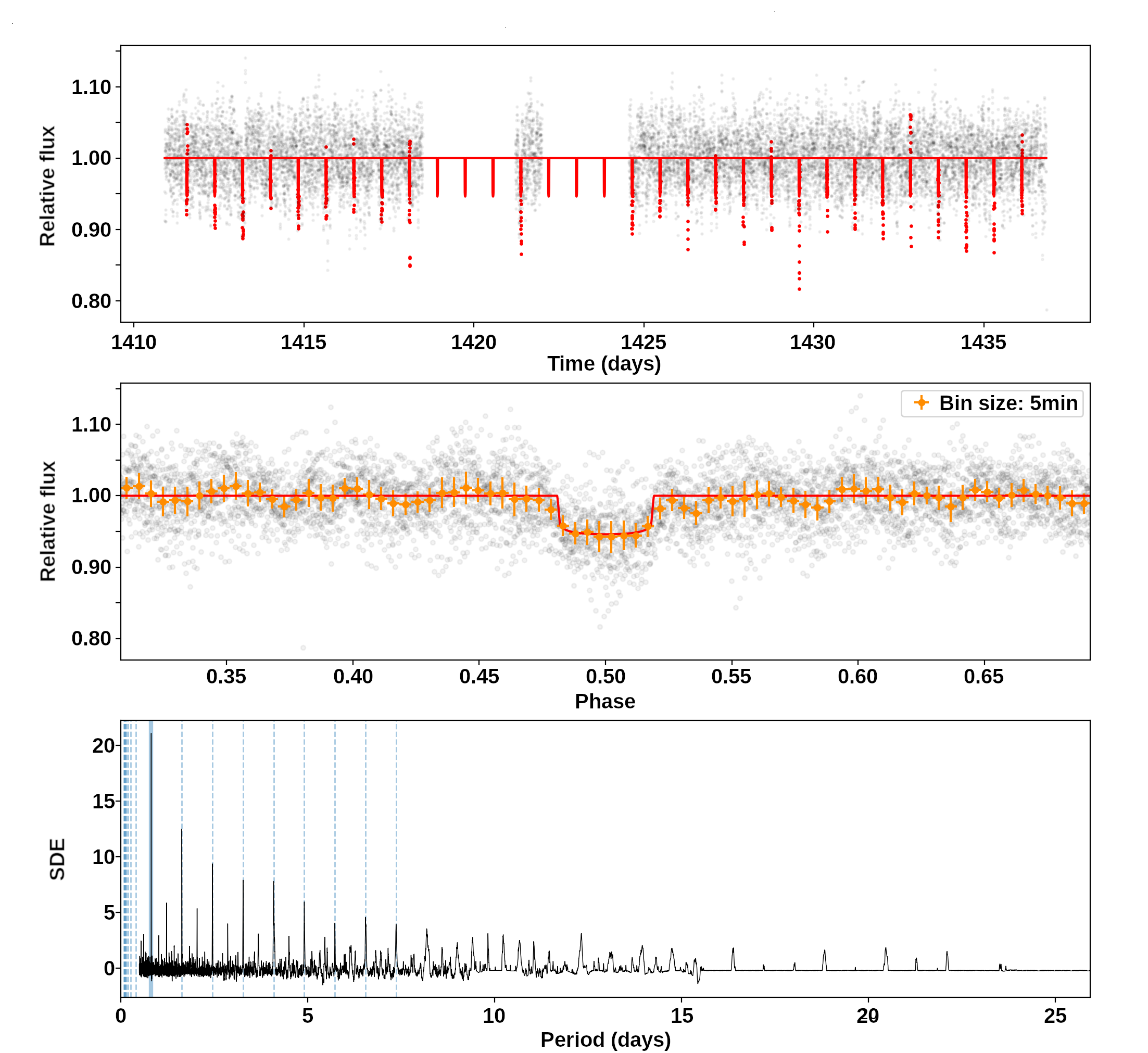}
        \includegraphics[width=9cm]{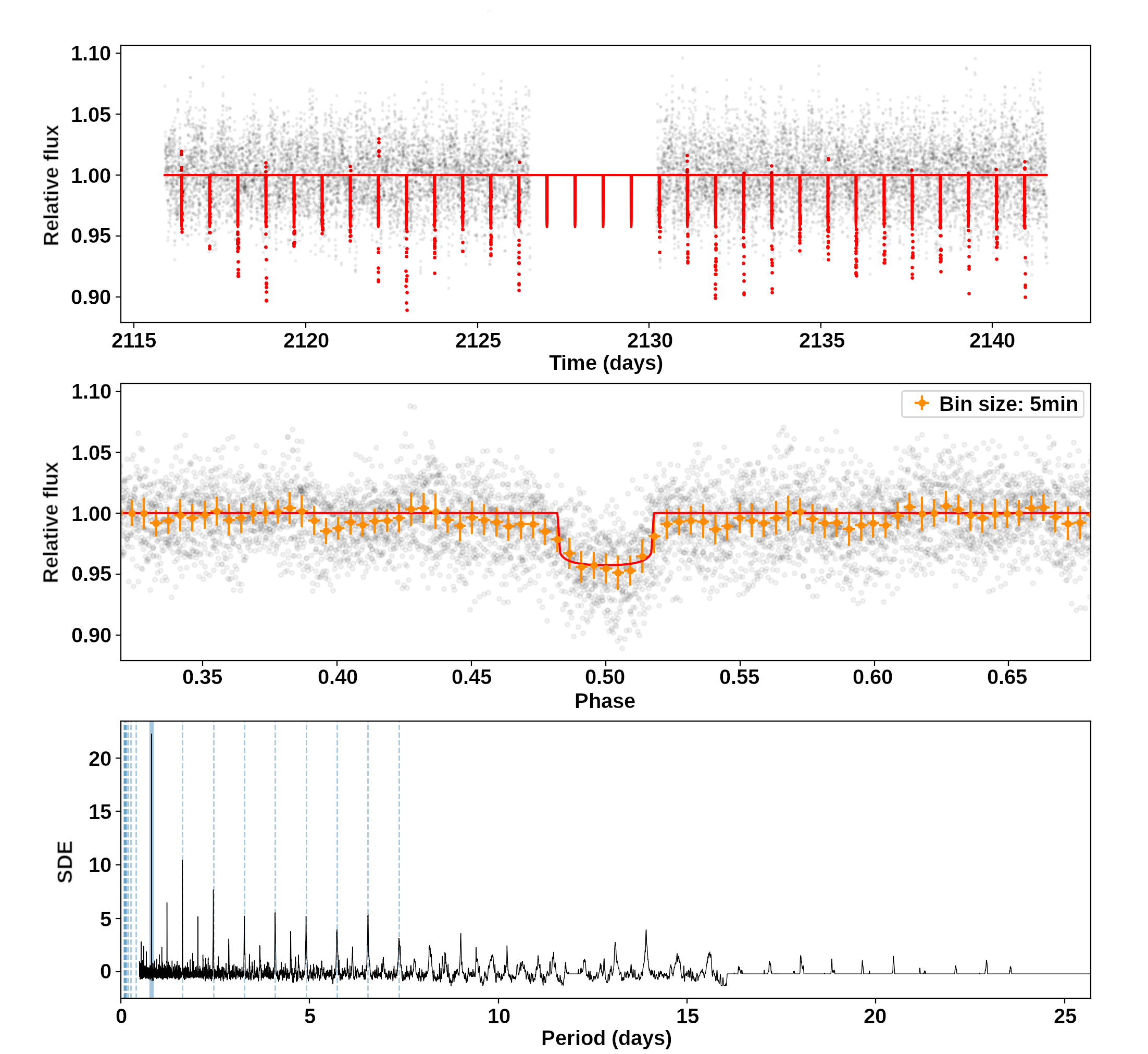}
        \centering
        \caption{\label{fig:TIC_398733009_TOI464.01} Detection of an interesting transit signal in the    light curve of TIC 398733009 in sectors 4 (three top panels) and 30 (three bottom panels).}
    \end{figure}

    Another star worth mentioning is TIC 398733009 (KUV 02062-0917). A very promising signal was detected in its light curve of sector 4, with a period of 0.82 days, and this signal is retrieved in sector 30 data (Fig. \ref{fig:TIC_398733009_TOI464.01}).
    
    However, further investigation gave contradictory insights. The transits are clear, but the eclipses (secondary transits) are also detected, which favours the explanation of a self-luminous body as a white dwarf. However, the computation of the equilibrium temperature of a body that close to its host star ($\sim$0.013 AU), following the approach described in \citet[][supplementary information, section D]{2011Natur.480..496C} gave a result between 5000 and 6000 K in its dayside, assuming an albedo of 0.1 and a heat redistribution coefficient $\beta$ between 0.2 and 1 (along with the canonical parameters for the sdB star $M_{\star} = 0.47 M_\odot$, $R_{\star} = 0.18 R_\odot$, and $T_{\rm eff} = 31000$ K). This heating could explain the secondary transits by itself, as such a hot planet emits its own light in the visible part of the spectrum, while the detection of both the transits and eclipses and their regularity (as they are separated by half of the orbital period) suggests a low eccentricity and an inclination close to 90$\circ$ degrees for the potential planet. The vetting phase supported an on-target origin for the transits and the absence of contamination from nearby stars, but TRICERATOPS gave an abnormally high false-positive probability (at 0.99 compared to only 0.45 for a non-false-positive probability). We therefore looked for other types of data for this target and built a SED from magnitudes in several bands. This SED revealed an excess in the red part of the spectrum that is too high to be explained by a hot planet. Instead, two stellar components are required to explain the observed colours. TIC 398733009 is therefore an EB system for which the exact nature of the stellar components has to be determined, in particular from high-resolution spectroscopy with a high signal-to-noise ratio. As this star was confirmed to be no hot subdwarf, it was removed from our target list for this project.
    
\begin{figure*}[ht]
        \includegraphics[scale=0.4]{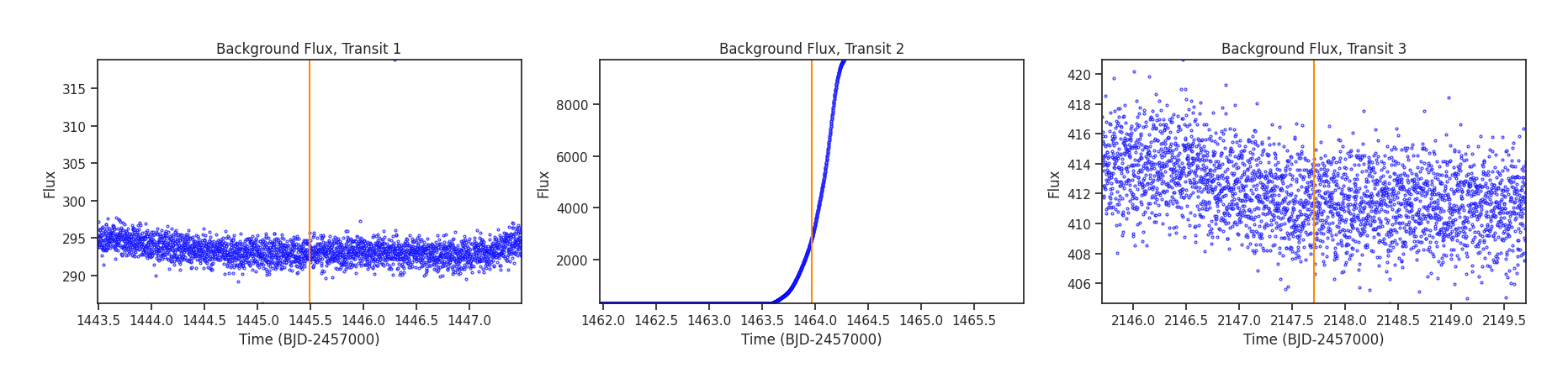}
        \centering
        \caption{\label{background} Background flux around the single transits for the detected signal in TIC 369394241 Sector 5. The transits' position are indicated by orange vertical lines. Left and right panels show low and constant values of background flux for the first and third transits. Centre panel shows a steep increase of background flux half a day before the second transit, reaching values 20 times higher than the baseline found for the other transits.}
    \end{figure*}

    One last example is the light curve of TIC 369394241 (HE0452-3654) that was observed in sector 5, in which we detected a promising signal that was retrieved in sectors 31 and 32. At stage 2 of the analysis (vetting and validation), however, we discovered that one of the transits occurred at a time when the background flux was very high above the acceptable thresholds, with a value of several thousands instead of the expected few hundreds (Fig. \ref{background}). In this case, the procedure was to mask the transit and repeat the analysis, in which we did not recover the signal. We later discovered that after an update of the SPOC algorithms from the TESS team, the portion corresponding to this transit is now masked in the most recent version of the PDC-SAP light curves. The second transit points are now flagged with quality bits 5 (Argabrightening Event) and 12 (Straylight). It was therefore labelled as a negative result (no signal) from this point.

\section{Discussion}
    \label{discussion}
    We have two main goals in this project. The first goal is to compute the occurrence rates of planets around hot subdwarfs, and the second goal is to provide observational constraints on the survival of close-in planets that are engulfed during the RGB phase of their host star.
    
    Concerning the first goal, we are already in the position to draw the very first statistics from the absence of a transit signal in 549 of our targets from TESS cycle 1. This non-detection scenario only allows us to compute the upper limit of the occurrence rates. We used the same method as in \citet{2018MNRAS.474.4603V} (sections 3.3 and 3.4), itself adapted from \citet{Faedi2011}, where this limit is written as
    \begin{equation}
        f_{max} = 1 - (1 - C)^{\frac{1}{N' + 1}}\;\;\;\;\;,
    \end{equation}
    with $f_{max}$ the upper limit for the occurrence rate of planets and $C$ the confidence level (between 0 and 1). $N'$ is defined as the product of $N$, the number of targets in the sample (549) by $P_{\rm transit}$, the geometrical transit probability, and $P_{\rm detection}$, the probability of detecting a transiting body. The probability $P_{\rm transit}$ is directly derived from parameters of the system by
    \begin{equation}
        P_{\rm transit} = \left(\frac{R_{\star} + R_{p}}{a}\right) \frac{1 + e \sin(\omega)}{1 - e^2}\;\;\;\;\;,
    \end{equation}
    where $R_{\star}$ and $R_{p}$ are the radii of the star and the planet, $a$ is the semi-major axis of the planet, $e$ is the eccentricity, and $\omega$ is the argument of the periapsis. We assumed circular orbits ($e=0$), inducing that the right fraction is equal to $1$. The only parameters that matter are therefore $R_{star}$, $R_{planet}$ , and $a$. This last parameter is recovered from the period of the planet via Kepler's third law \citep{Kepler}. Stellar properties are assumed equal to canonical parameters for sdB stars with $R_{\star} = 0.17 R_{\odot}$\footnote{Canonical radii for hot subdwarf stars are often chosen to be 0.17 and 0.18 $R_{\odot}$ depending on the team or scientist doing the computation, hence the difference this value and the value used in Sect. 3. The resulting difference is negligible in our case.} and $M_{\star} = 0.47 M_{\odot}$. As hot subdwarfs are small stars, we cannot neglect planetary radii, hence the equation becomes
    \begin{equation}
        P_{\rm transit} = \frac{R_{\star} + R_{p}}{a}
    .\end{equation}
    
    The probability $P_{\rm detection}$ is computed using injection-and-recovery tests for various periods, $R_{p}$ , and canonical parameters for the host stars within pre-determined magnitude bins. Many of these injection-and-recovery tests were performed in \citet{Van-Grootel-21}, from which we found that our ability to detect planets in light curves provided by the space telescopes primarily depends on the magnitude of the star (see their Table 3 and Fig. 3, in particular).
    
\begin{figure}[ht]
        \includegraphics[width=9cm]{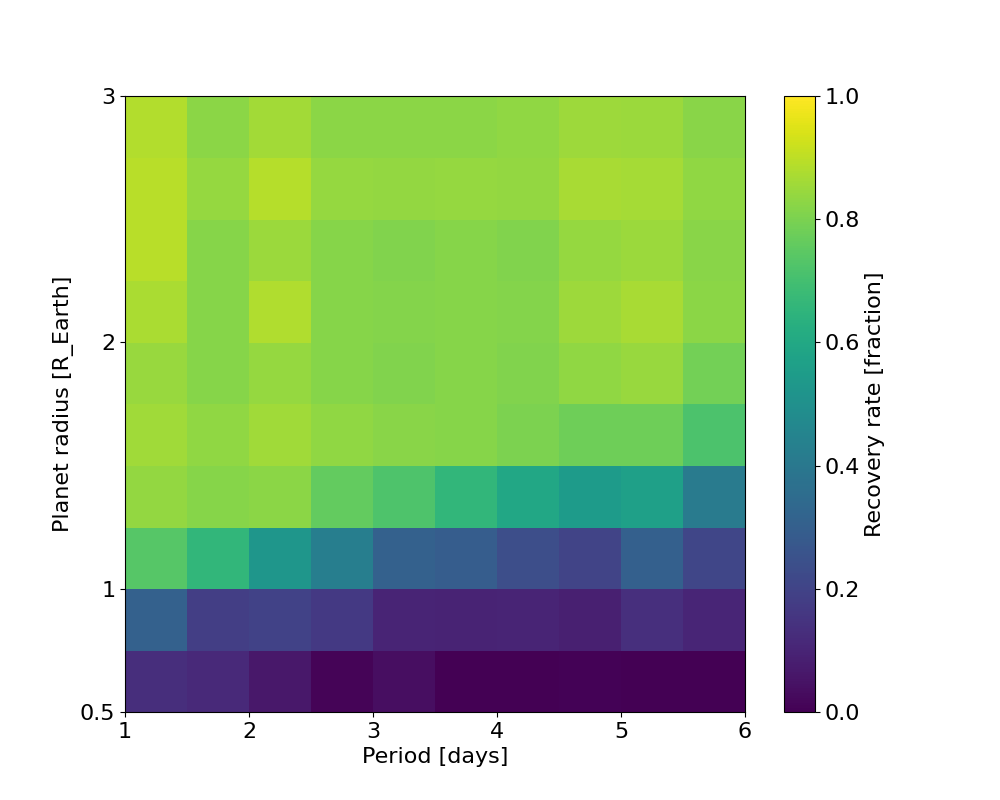}
        \includegraphics[width=9cm]{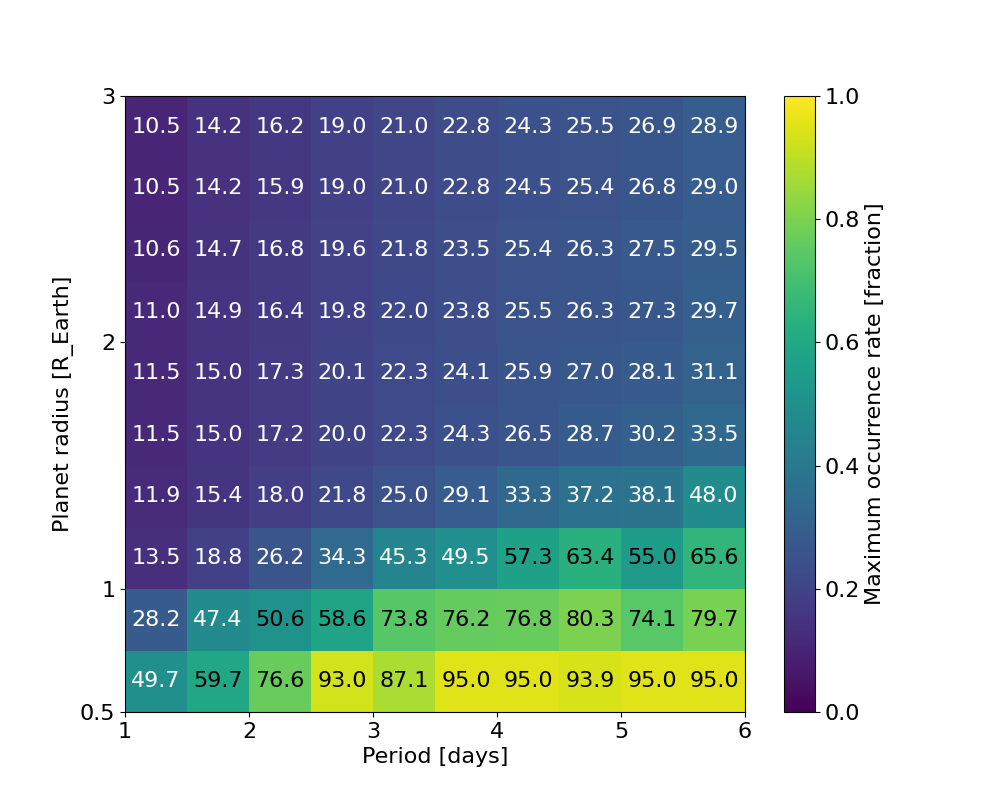}
        \centering
        \caption{\label{fig:Occurrences} Upper limit $f_{max}$ for the occurrence rate of planets around hot subdwarfs from our sample of 549 stars without a detection, considering they have a G magnitude between 13 and 13.5. The two panels share the same axes with the period (in days) in abscissa and radius of the orbiting body (in $R_{\oplus}$) in ordinate. We focus here on short-period planets (1 to 6 days) of few Earth radii (0.5 to 3). \textit{Top panel}: Recovery rate $P_{detection}$ (deep blue = 0, yellow = 1). \textit{Bottom panel}: Upper limit for the occurrence rate $f_{max}$ (deep blue = 0, yellow = 1). For this graph, the confidence level was set to 0.95. For readability, the value of the limit is also written in each cell.}
    \end{figure}

    Figure \ref{fig:Occurrences} presents the very first estimate for the occurrence rates of planets around hot subdwarfs, making the simplification that all the 549 stars have G magnitude between 13 and 13.5\footnote{Weaker constraints are expected for fainter stars and stronger constraints for brighter stars. Stars with magnitudes between 13 and 13.5 are in the brighter part of our sample (see \citealt{Van-Grootel-21}), but were selected for this very first estimate because they are the most numerous in the injection-and-recovery tests in our hands, namely, those performed for \cite{Van-Grootel-21}. Each computation indeed requires thousands of injections and recoveries and can take weeks, therefore we relied on those already in our possession.}. When our sample is larger and all magnitude bins are more populated, occurrence rates will be weighted according to the numbers of stars in each magnitude bin. We will also consider actual stellar radii.

    The top panel of Fig. \ref{fig:Occurrences} focusses on the injection-and-recovery tests, and the bottom panel displays the corresponding upper limit $f_{max}$ for the occurrence of planets for each period-radius pair. For example, at a one-day orbital period, we can exclude the presence of 3$R_{\oplus}$ planets in 90\% of hot subdwarfs, and the presence of 0.5$R_{\oplus}$ planets in 50\% of them. At 6d, these numbers decrease to 71\% (3$R_{\oplus}$) and 5\% (0.5$R_{\oplus}$). As expected, the constraint is less stringent when the detection capabilities are lower, which is shown by the higher values for the upper limit for smaller planets and longer periods.
    
    The second goal of our project is to place observational constraints on the survival of close-in planets that are engulfed during the RGB phase of their host star. We wish to list some considerations here. Firstly, to achieve this objective, we restrict our sample to sdB stars, which are thought to be in their vast majority direct post-RGB stars, the evolutionary paths of sdO stars are thought to be more diverse as some may not be direct post-RGB stars (see Sect. \ref{Intro}). Secondly, given the nature of the transit method, a detection will likely correspond to a close-orbiting planet. Three scenarios could explain this configuration: it could be (1) a non-engulfed, far-orbiting planet that migrated inwards after the end of the RGB phase, (2) a second-generation planet formed using remnant material present in the system after the ejection of the RGB envelope, and (3) a planet that survived the engulfment, at least partially.
    
    Although dedicated computations are required, the migration of bodies at greater distances that were not engulfed in the envelope of the red giant star is difficult because sdB stars represent a short phase of stellar evolution ($\sim$150 Myr for the core-He burning, i.e., EHB, phase; \citealt{2016PASP..128h2001H}) and planetary migration from the outer part of the system is a long process \citep{2018MNRAS.476.3939M}. For the same reason, the formation of second-generation planets is unlikely, in particular in light of the harsh environment for planet formation around a hot subdwarf. The survival of an engulfed planet is thought to be very difficult, but is completely unconstrained from observations. Furthermore, not only does the small hot subdwarf size enable the detection of small, possibly disintegrating remnant objects, but the ejection of the red giant envelope, which is necessary to form a hot subdwarf, may even be the reason for the survival of such remnants by stopping the spiralling-in inside the host star.
    
    
    Computing the number of expected detections is a difficult exercise. Simulations from \cite{Staff2016} indicate a quick spiralling of giant planets onto the star. However, some relatively low-mass brown dwarfs are known to be very close companions to sdB stars \citep{Schaffenroth2014, Schaffenroth2021} and hence have survived this engulfment, so that we can imagine that massive planets might survive engulfment in their host star as well. Therefore, if any, the most probable bodies able to survive would be the cores of former giant planets, whose atmospheres would have been stripped out. The exact nature of these cores is thought to be a few Earth radii with a dozen Earth masses in the currently dominant core-accretion model \citep{1996Icar..124...62P}. However, we note that  the presence of this core is still uncertain even for Jupiter \citep{2017GeoRL..44.4649W}. For instance, the occurrence rate of hot-Jupiter planets around Sun-like stars, which are hot subdwarf progenitors, is approximately 1\% \citep{Wright-12}. In the hypothetical case in which all their cores survive, there should be 23 to 50 (for 2300 targets in SC to 5000 targets at all cadences; \citealt{Van-Grootel-21}) close-orbiting remnants in our sample. Accounting for the transit detection probabilities, there should be about two to five detections at orbital periods shorter than 10 days in these 2300 to 5000 targets. However, this only includes hot-Jupiter planets. Other types of planets may also survive.
 
    In an upcoming part of the project we will also search for disintegrating planets, such as debris and/or planets with dust tails, with a new module that we are building in the {\fontfamily{pcr}\selectfont SHERLOCK} pipeline. This upgrade will allow us to search for transits with asymmetric shapes that are produced by the dust tails, such as the one found around a white dwarf by \cite{2015Natur.526..546V}. One last point is the potential role that giant planets might have in the formation of hot subdwarfs in the case of an engulfment. If it is large enough, the planet might help to expel the envelope, but this possibility is debated (e.g. \citealt{2020A&A...642A..97K}). The detection of remnants around apparently single hot subdwarfs may provide new elements to settle this issue.

\begin{acknowledgements}
    We warmly thank the anonymous referee for constructive remarks that improved our paper. We thank Uli Heber and Elizabeth M. Green for their help on the characterisation of several of our targets, as well as attendees of the sdOB9.5 conference in Potsdam, namely but not limited to, Stephan Geier and Philipp Podsiadlowksi, as the discussion there were of great interest for this work.
    
    This work has been supported by the University of Liège through an ARC grant for Concerted Research Actions financed by the Wallonia-Brussels Federation. A.T. acknowledge financial support from the ULB "Fond de rattrapage PDR". V.V.G. and L.S are senior F.R.S.-FNRS Research Associates. SC acknowledges financial support from the Centre National d’Études Spatiales (CNES, France). This paper includes data collected by the TESS mission. Funding for the TESS mission is provided by the NASA Explorer Program. Funding for the TESS Asteroseismic Science Operations Centre is provided by the Danish National Research Foundation (Grant agreement no.: DNRF106), ESA PRODEX (PEA 4000119301) and Stellar Astrophysics Centre (SAC) at Aarhus University. We thank the TESS team and staff and TASC/TASOC for their support of the present work. This work has made use of data from the ESA mission Gaia (https://www.cosmos.esa.int/gaia), processed by the Gaia Data Processing and Analysis Consortium (DPAC, https://www.cosmos.esa.int/web/gaia/dpac/consortium). Funding for the DPAC has been provided by national institutions, in particular the institutions participating in the Gaia Multilateral Agreement.
\end{acknowledgements}

\bibliographystyle{aa}
\bibliography{Biblio/references}

\appendix
    \section{List of stars without planetary transit signals above thresholds}
        \begin{figure*}[h!]
            \label{figA1}
            \includegraphics[scale=0.5]{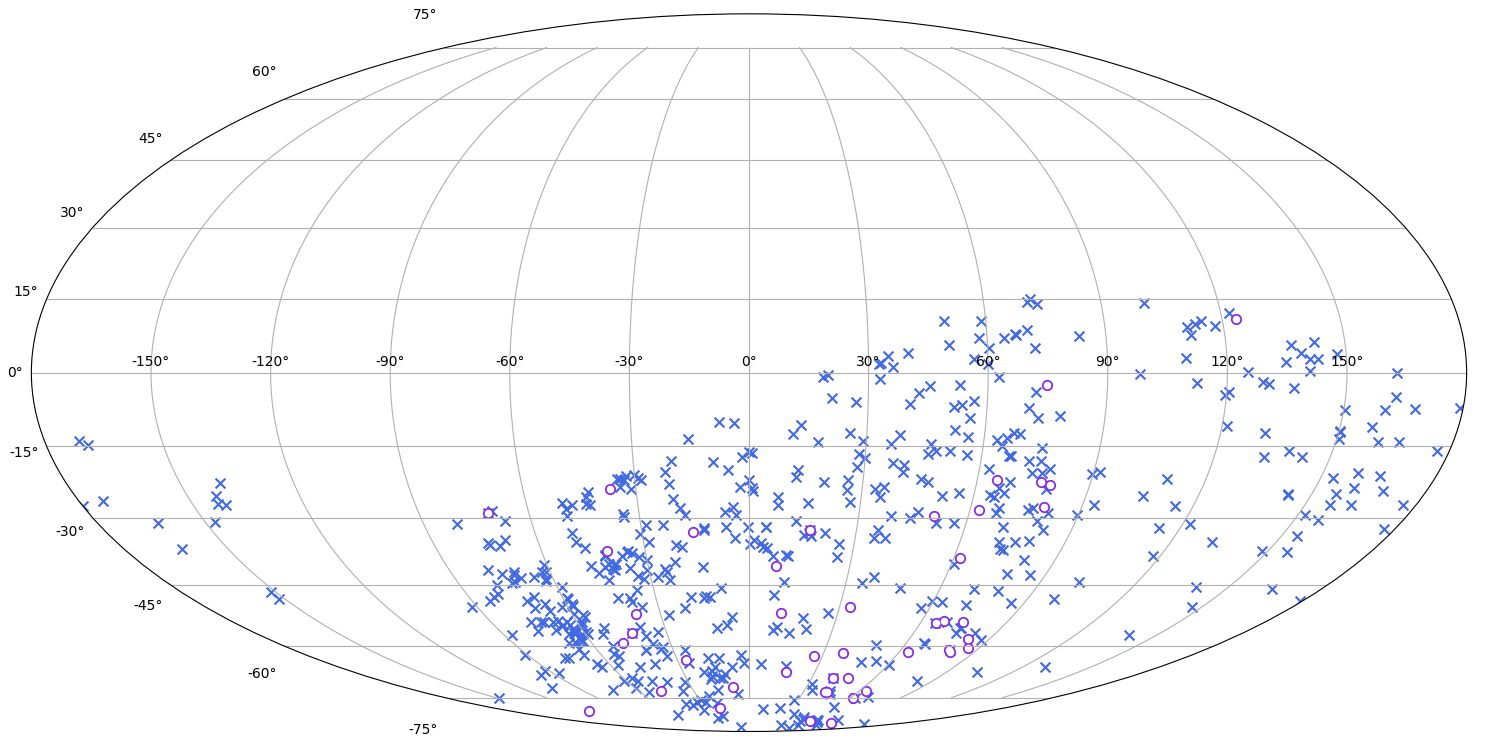}
            \centering
            \caption{Celestial distribution of the targets without a detected signal (blue crosses) and those that reached stage 3 of the analysis and for which follow up is scheduled (open purple circles). See also Fig. 1 from \cite{Van-Grootel-21} for the global distribution of hot subdwarf stars observed by the missions Kepler, K2, TESS, and CHEOPS. Data shown in equatorial coordinates and in a Mollweide projection of the sky.}
        \end{figure*}
        
        \begin{table*}[pt]
        \centering
            \caption{\label{Negative_list} List of targets (ranked by increasing TIC number) without a detected planetary transit signal above our thresholds (they did not reach stage 1 of the analysis). See details in Sect. \ref{results}.}
                \begin{tabular}{|*{9}{r}|}
                    \hline\hline
                        8842 & 61379646 & 91556284 & 146437397 & 189585273 & 248949857 & 289795477 & 352480413 & 410135274 \\
                        1526470 & 61621625 & 92226759 & 146520657 & 193087619 & 253656464 & 289823802 & 354572272 & 410280069 \\
                        2213137 & 61728030 & 92984690 & 147115112 & 193092806 & 253932935 & 290401825 & 355002205 & 410390905 \\
                        3990402 & 62023018 & 94945019 & 147283842 & 197570382 & 254263628 & 290646079 & 360736086 & 413300076 \\
                        5051080 & 62223401 & 98868202 & 147349694 & 197614201 & 254287117 & 293165262 & 360804908 & 415339307 \\
                        6593243 & 62381958 & 98871628 & 148993614 & 197687846 & 255697796 & 293463617 & 360806208 & 419571221 \\
                        7319744 & 62483415 & 101403951 & 149767908 & 197693940 & 259392878 & 293520466 & 364424541 & 419998091 \\
                        9053429 & 63428034 & 101454745 & 149846788 & 197765610 & 259539930 & 293791715 & 369377398 & 420049852 \\
                        9102069 & 63695388 & 101512805 & 151892844 & 200388625 & 259864042 & 294836239 & 369965957 & 421895532 \\
                        9268360 & 63696810 & 101540375 & 152286180 & 200436339 & 260749705 & 300607115 & 371801053 & 421939894 \\
                        9358354 & 64008380 & 101545865 & 152373379 & 200545781 & 260795163 & 301022374 & 372172495 & 421951567 \\
                        10932480 & 64111698 & 101817287 & 152374958 & 201690280 & 260839766 & 301405970 & 373245897 & 421999342 \\
                        12378718 & 64112207 & 101817673 & 153165824 & 206347723 & 261241692 & 301454053 & 374408958 & 422004014 \\
                        12528447 & 65145453 & 107548305 & 155776755 & 206462125 & 261262954 & 304103779 & 379352979 & 422043417 \\
                        12549312 & 65324889 & 111184069 & 156215534 & 206469610 & 261380566 & 307934757 & 380274928 & 422149668 \\
                        13069774 & 66398320 & 113118441 & 157323544 & 206482194 & 261427146 & 309754884 & 380590152 & 423160573 \\
                        13145616 & 66493797 & 114115352 & 158235404 & 206590489 & 261679852 & 317058444 & 382247276 & 423761655 \\
                        20448010 & 66626624 & 114502553 & 158335560 & 207208668 & 262039948 & 317154662 & 382375326 & 423763742 \\
                        24448566 & 67583991 & 114807149 & 158343496 & 207328336 & 262818577 & 317350896 & 382383606 & 424881855 \\
                        24908871 & 67584818 & 115175941 & 159640970 & 209393544 & 262921847 & 317439554 & 382415345 & 424941595 \\
                        25136499 & 67598107 & 115273584 & 159669717 & 209397773 & 262927968 & 319602897 & 382416576 & 425064757 \\
                        25137423 & 69841801 & 115692439 & 159693368 & 212320065 & 265445890 & 320092330 & 383217734 & 425206178 \\
                        25245570 & 69911593 & 117626475 & 159747831 & 212347311 & 266680031 & 320173712 & 386642921 & 425516683 \\
                        25300088 & 70723238 & 117735611 & 159805154 & 212352349 & 269762836 & 320176500 & 386653873 & 425799931 \\
                        25948892 & 70776745 & 118327563 & 159833896 & 214568914 & 269855226 & 320417198 & 386703105 & 426030266 \\
                        28052378 & 70962870 & 118412596 & 159843438 & 218960493 & 269973828 & 320484212 & 387107334 & 432162081 \\
                        29821374 & 71013467 & 118588732 & 160078278 & 218961596 & 270000741 & 320529836 & 389476185 & 432744391 \\
                        29840077 & 71109189 & 120580052 & 160797304 & 219225205 & 270071350 & 320533176 & 389520459 & 436579904 \\
                        30019744 & 71133157 & 120610731 & 161153327 & 219974863 & 270245694 & 320591159 & 389752750 & 436636292 \\
                        30415110 & 71150825 & 121318590 & 161402643 & 220026025 & 270285517 & 320591947 & 392703299 & 436682542 \\
                        31174073 & 71248239 & 121550523 & 164754858 & 220347928 & 270380474 & 320660807 & 392758248 & 437237493 \\
                        32354769 & 71345281 & 122521574 & 165110581 & 220370211 & 270394275 & 320939631 & 393033236 & 439461184 \\
                        32661254 & 71410075 & 123665218 & 166395799 & 220472655 & 270416019 & 320951753 & 393491149 & 439905042 \\
                        33318760 & 71716888 & 124029363 & 166748749 & 220476769 & 270562073 & 320965274 & 393941149 & 441399312 \\
                        33321190 & 72667488 & 124175842 & 167456967 & 220479184 & 271576664 & 322285377 & 394018151 & 441401311 \\
                        33490778 & 72763826 & 125556577 & 167746025 & 220573709 & 273218137 & 325394866 & 394494920 & 441499030 \\
                        33526769 & 73185296 & 126637970 & 167976324 & 228508601 & 273862178 & 325566833 & 394496912 & 441508514 \\
                        33767134 & 73764693 & 126803779 & 168304840 & 229050493 & 274035031 & 326096162 & 394517648 & 443625969 \\
                        33944704 & 76184341 & 128971400 & 169285097 & 229051528 & 275185985 & 326328277 & 394517681 & 445927286 \\
                        36729387 & 77005197 & 131343095 & 170108378 & 229144939 & 275921038 & 326453105 & 394678374 & 455206965 \\
                        36995993 & 77360048 & 138707463 & 170203297 & 231308507 & 277773221 & 328179553 & 394698511 & 457168745 \\
                        37602756 & 77477081 & 138707823 & 170869314 & 231629787 & 277892210 & 332697630 & 396004353 & 457196370 \\
                        38423413 & 77959296 & 139208432 & 176380024 & 231695087 & 278659026 & 332701732 & 396695965 & 459932445 \\
                        38511369 & 77992461 & 139266474 & 177079697 & 231712886 & 278705842 & 332742020 & 398796167 & 464549287 \\
                        44625421 & 79246796 & 139682931 & 178345419 & 231812407 & 279342801 & 333659638 & 398801916 & 464769756 \\
                        45362220 & 79493362 & 139804925 & 178873605 & 234281664 & 280051980 & 339262037 & 403317759 & 466229760 \\
                        47352092 & 79956635 & 140494440 & 178893906 & 234295068 & 280775141 & 339381798 & 404158419 & 466273074 \\
                        47355413 & 79958453 & 142200764 & 178896096 & 234386436 & 280789378 & 339907982 & 404430996 & 466310972 \\
                        47377536 & 80057233 & 142875987 & 179034615 & 234526947 & 281595262 & 341001113 & 404467239 & 469791892 \\
                        47482655 & 80170223 & 143058705 & 179278778 & 235011371 & 281658096 & 343828974 & 405799245 & 469981019 \\
                        49711130 & 80290366 & 143699381 & 181820016 & 237322080 & 281851153 & 345449417 & 406239686 & 471015202 \\
                        50384080 & 80427831 & 143923307 & 181914779 & 237338096 & 283866221 & 347412256 & 406241063 & 471015203 \\
                        53939726 & 83755080 & 144193687 & 182012072 & 238853890 & 284677903 & 349367583 & 406242857 & 471015416 \\
                        54986420 & 88417452 & 144306296 & 183530773 & 241675531 & 284703017 & 350155206 & 406280054 & \\
                        56124677 & 89148712 & 144804862 & 183563651 & 246881248 & 286099192 & 350273432 & 406280906 & \\
                        56648314 & 89467049 & 146251617 & 186149538 & 246881770 & 289533626 & 350583903 & 407768021 & \\
                        59769766 & 89529774 & 146282446 & 188174892 & 248382748 & 289721074 & 352142391 & 409644971 & \\
                        61029108 & 91316983 & 146323153 & 189585096 & 248391428 & 289737935 & 352315023 & 409850857 & \\
                    
                    \hline \hline
                \end{tabular}
        \end{table*}


\end{document}